\documentclass[10pt,conference]{IEEEtran}
\usepackage{cite}
\usepackage{color}
\usepackage[stable]{footmisc}
\usepackage{siunitx}
\usepackage{authblk}
\usepackage[center]{subfigure}
\usepackage{verbatim}
\usepackage[super]{nth}
\usepackage{comment}
\usepackage{afterpage,lipsum}

\usepackage{amsfonts}
\usepackage[colorinlistoftodos]{todonotes}
\usepackage[linesnumbered, lined, commentsnumbered, ruled,vlined, longend, noend]{algorithm2e}
\usepackage{algpseudocode}
\usepackage{mathtools}
\usepackage{dirtytalk}
\usepackage{float}
\usepackage{times,amsmath,epsfig,amssymb,cite,bm,float,epstopdf,graphicx,graphics,amsthm}
\usepackage{forloop}
\usepackage{enumitem}
\usepackage{caption}
\usepackage{multicol}
\newcounter{ct}

\newcommand{\cmmnt}[1]{\ignorespaces}

\usepackage{fancyhdr,lipsum}
\fancypagestyle{mahmood}{%
   \fancyhf{} 
   
   \fancyhead[C]{Accepted paper for publication in Globecom 2025. You can use this material personally. Reprinting or republishing this material for the purpose of advertising or promotion, creating new collective works, reselling or redistributing to servers or lists, or using any copyrighted component in other works must adhere to IEEE policy. The DOI will be supplied as soon as it becomes available.}

}%
\makeatletter
\let\ps@IEEEtitlepagestyle\ps@mahmood
\makeatother
\begin{document}

\title {Deep Learning based Moving Target Defence for Federated Learning against Poisoning Attack in MEC Systems with a 6G Wireless Model \vspace{-0.2cm}}

\author{Somayeh Kianpisheh\textsuperscript{1}}
\author{Tarik Taleb\textsuperscript{2}}
\author{Jari Iinatti\textsuperscript{1}}
\author{JaeSeung Song\textsuperscript{3}}

\vspace{-6cm}
\affil{\textsuperscript{1}\textit{Centre for Wireless Communications, University of Oulu, Finland}, \textsuperscript{2} \textit{Ruhr University Bochum, Germany} \\

\textsuperscript{3} \textit{Computer and Information Security Department, Sejong University, South Korea}
\\
\textit{Emails: somayeh.kianpisheh@oulu.fi, tarik.taleb@rub.de, jari.iinatti@oulu.fi, jssong@sejong.ac.kr \vspace{-0.6cm}}}

\maketitle
\begin{abstract}
Collaboration opportunities for devices are facilitated with Federated Learning (FL). Edge computing facilitates aggregation at edge and reduces latency. To deal with model poisoning attacks, model-based outlier detection mechanisms may not operate efficiently with hetereogenous models or in recognition of complex attacks. This paper fosters the defense line against model poisoning attack by exploiting device-level traffic analysis to anticipate the reliability of participants. FL is empowered with a topology mutation strategy, as a Moving Target Defence (MTD) strategy to dynamically change the participants in learning. Based on the adoption of recurrent neural networks for time-series analysis of traffic and a 6G wireless model, optimization framework for MTD strategy is given. A deep reinforcement mechanism is provided to optimize topology mutation in adaption with the anticipated Byzantine status of devices and the communication channel capabilities at devices. For a DDoS attack detection application and under Botnet attack at devices level, results illustrate acceptable malicious models exclusion and improvement in recognition time and accuracy.   
\end{abstract}

\IEEEpeerreviewmaketitle
\textbf{Index Terms-} federated learning, poisoning attack, MEC, 6G, moving target defence, deep reinforcement learning.
\section{INTRODUCTION}
The collaboration opportunities for smart devices are facilitated with Federated Learning (FL) by training a global model from distributed data while maintaining data privacy preservation by sharing only model parameters \cite{mcmahan2017communication}. At each iteration of the process, locally trained models are transmitted to an aggregator server to construct a global model which evolves within iterations. To overcome the issues of intolerable latency in a centralized server scenario, exploiting edge computing is an alternative \cite{zhou2022differentially}. Multi-Access Edge Computing (MEC) capabilities at base stations, can be exploited either for the aggregation at BSs for a fast response, or for partially aggregation of the parameters and transfer them to a server for the global aggregation \cite{zhou2022differentially}. Most FL studies in edge computing assume a secure FL protocol. 

A model poisoning attack exploits system vulnerabilities and injects poisoned local model updates, causing a useless or less accurate global model. To make the FL robust, defense mechanisms mainly focus on mitigating model poisoning attack through outlier detection mechanisms. The studies in \cite{guerraoui2018hidden}, \cite{yin2018Byzantine} advocate robust aggregation rules that detect outlier models and remove them in the aggregation. The study in \cite{chen2021fedequal} emphasizes on the performance of the learning process in an outlier detection application. Cost-efficient method to isolate the detected attackers has been presented in \cite{huang2021cost}. The studies in \cite{al2023untargeted}, \cite{pan2020justinian} find unreliable models by analyzing the behaviour of the clients from the aspect of model performance. Accordingly they remove them in the aggregation phase. The robustness of FL against Byzantine clients while preserving privacy has been investigated in \cite{Jinhyun2024Byzantine}. 

There are two drawbacks for the mentioned works: First, the differentiation of malicious from benign model might not be efficient in heterogeneous local modes, where devices have driven the models based on part of the data \cite{chen2021fedequal}. This is particularly observable at early iterations when there is no knowledge about the global model \cite{chen2021fedequal}. Second, outlier detection-based mechanisms might not be efficient in the recognition of attacks with complex patterns or behaviors \cite{baruch2019little}. To overcome the mentioned drawbacks, this paper fosters the defense line in two ways: First, inspired by the fact that the attacker sends traffic toward a device to perform malicious tasks \cite{zhang2023moving}, \cite{sharafaldin2018toward}, \cite{kianpisheh2024multi} (e.g., discover device vulnerabilities, to get control and update poisoned models), this paper uses device-level traffic analysis to estimate the reliability of participants in learning. This estimation can particularly be utilized when model-based outlier detection may not be efficient. Second, in contrast to the reactive approach in the literature, a proactive approach will be applied to reduce the opportunity for poisoning. The study in \cite{kianpisheh2024multi} reduces the opportunity of poisoning by fooling the attacker by employing several models to train the main model. However, the method is not applicable if there is full control over the training process on compromised devices.    

The proactive Moving Target Defence (MTD) concept reduces cyber-threats by dynamically adjusting network properties to distort adversary knowledge to trigger the attack \cite{zhang2023moving}. Inspired by the idea of MTD, this paper introduces \textbf{MTD-}based \textbf{FL} (MTD-FL) which empowers FL with a topology mutation strategy, as an MTD strategy to distort the assumption of participation of all devices in FL. Based on device-level traffic analysis, a topology mutation strategy dynamically changes participants to remove the devices that can be targeted in the future.   

Recurrent Neural Network (RNN) which has shown efficiency in time-series analysis of security applications \cite{assis2021gru}, \cite{kianpisheh2024multi}, is adopted for device-level traffic analysis and poisoning-occurence prediction. Optimization framework is provided for topology-mutation based MTD scheme. To provide optimal MTD strategy in large state/action spaces and under dynamic nature of wireless communication channels and devices mobility, a Deep-Reinforcement Learning (DRL) based mechanism is proposed. For a Distributed Denial of Service (DDoS) attack detection scenario, the results illustrate an improvement in accuracy and recognition time.

\section{SYSTEM MODEL}
\textbf{Network:} The network has $N$ mobile IoT devices, $M$ Base Stations (BSs) equipped with MEC processing, and a cloud. The CPU frequency at $i^{th}$ $BS$ i.e., $BS_i$ and the cloud are $f^{cmp}_i$ and $f^{cmp}_c$, respectively. CPU frequency of the device $u$ is $f^{cmp}_u$. $R_{i,c}$ is the bandwidth of communication between BS $i$ and the central cloud. $R_i$ is the devices under coverage of the base station $i$. $B_i$ is transmission bandwidth of the base station $i$ to communicate with devices. The device $u$ has data of size $|D_{u}|$ and denoted by $D_{u}=\{(x^{u}_{1},y^{u}_{1}),... (x^{u}_{|D_{u}|},y^{u}_{|D_{u}|})\}$; where, $y$ is the label for input $x$. \\
A global model is constructed by performing FL and data sharing through MEC/cloud infrastructure. Using local data in the devices, the global model $M_c$ is trained in iterations. A device trains a local model at each iteration, based on which the global model is evolved through the learning process and the required recognition is performed.   
 
\textbf{Adversary Knowledge and Operation:}  
At any iteration, the adversary can exploit devices' vulnerabilities and compromise them to inject poisoned models in learning process. The compromised devices are Byzantine devices. The attacker does not have any control over the aggregation process at MEC, nor over the protocol of the benign devices which follow a normal implementation of the protocol.

Adversary establishes poisoned model(s) through applying an algorithm and uploads the malicious model(s) on behalf of the compromised device(s) when communicating with MEC for aggregation \cite{al2023untargeted}, \cite{kianpisheh2024multi}. This paper's MTD-based FL scheme can be applied for either targeted or untargeted attacks. In targeted attacks, malicious models are crafted so that the trained global model approaches a targeted model \cite{al2023untargeted}. Otherwise, an untargeted deviation strategy is applied \cite{al2023untargeted}.

\textbf{Defender Objectives:} MTD-FL scheme aims to train global model with applying MTD strategy through iterations to make the FL robust with a predefined confidence level, while keeping the performance of FL in terms of accuracy and recognition time, which is critical in ultra-low latency applications.

\begin{figure}
\begin{center}
    \includegraphics[width=8.2cm, height=5cm]{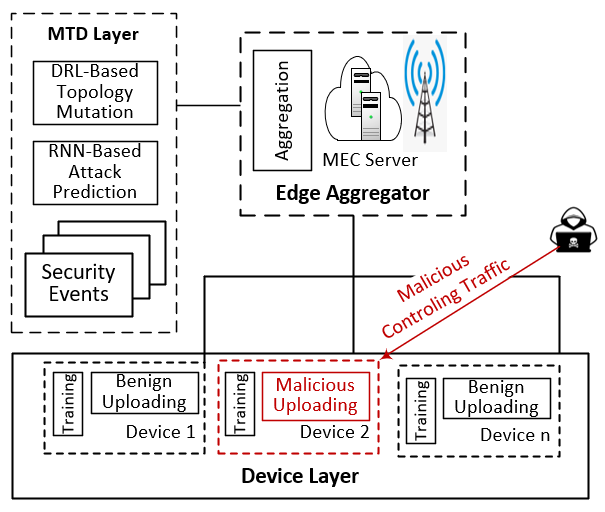}
    \vspace{-0.2cm}
    \caption{MTD-FL scheme.}
    \vspace{-1cm}
\label{fig:arch}
\end{center}
\end{figure}

\section{Moving Target Defence Based FL Scheme}
The MTD-FL scheme  overview, the RNN-based traffic analysis and the optimization are discussed in this section.    
\vspace{-0.5cm}
\subsection{Optimization Motivation and Scheme Overview}
The participants in FL define the learning topology. The MTD-FL scheme mutates the topology as an MTD strategy, to make it harder for an attacker to maintain a foothold. The topology mutation is performed in response to suspicious activities in learning iterations. Fig. \ref{fig:arch} shows the MTD-FL scheme. The histories of security events of the devices are logged. A RNN-based scheme is employed to predict the poisoning attack threats in the devices. Based on the predicted values of RNN and the state of the network, the state of the system is determined. Accordingly, a DRL based scheme is employed for topology mutation strategy, based on which the participants in the next iteration are specified. As an example, in Fig. \ref{fig:arch}, when an attacker targets device 2, it sends malicious controlling traffic to device 2 to control and perform malicious uploading. The RNN model that has been trained with traffic patterns in the security events of device 2, recognizes the arrived traffic as abnormal and anticipates that device 2 may hold a malicious model in the next iterations. Then the FL topology is adjusted to exclude devices 2. 
 
The optimization of topology mutation demands decision about participation of devices in federation. The optimal decision for inclusion of a benign-anticipated device (with high confidence) in learning depends on the trade-off between accuracy and time. Since exclusion may reduce the FL time, particularly if the device has low processing speed or if the device is straggler with poor channel condition to communicate with edge. However, the accuracy might be reduced due the missing of the device knowledge; thereby the necessity of an appropriate trade-off and optimization. Furthermore, the dynamicity of the communication channel makes the optimization challenging. The optimal decision for exclusion of a Byzantine-anticipated device depends on the confidence of prediction. For high-confidence exclusion can increase the chance of poisoning threat mitigation. However, for middle-range confidence (e.g., 40\% till 60\%), the accuracy might be reduced by exclusion of device if the device is benign, while there is the potential for attack mitigation if the device is poisoned. Thus, the problem of devices selection introduces an optimization to properly do the required trade-off among time and accuracy while ensuring a confidence for poisoning threat mitigation. Based on devices' reliability assessments, performance of FL and wireless communication channel, the topology mutation is optimized. 

\vspace{-0.1cm}
\subsection{RNN based Model Poisoning Attack Prediction}
Since the adversary sends malicious traffic to compromise a device and gain control over the upload of the model, analyzing traffic on devices to detect abnormal patterns can be a clue in the prediction of the target devices. Through monitoring, security event sequences on devices is analyzed to predict target devices in future. Since RNNs e.g., Gated Recurrent Units (GRU), Long Short-Term Memory (LSTM) have shown to be promising in security related analysis \cite{kianpisheh2024collaborative}, \cite{feltus2022learning}, we also adopt them for poisoning attack prediction. Let $E_u=\{e_{1}^u,..., e_{t}^u\}$ be the log collected within time-series to reflect poisoning-related security events in device $u$. The prediction model obtained from neural network is represented with $\phi(X_u; \theta_u)$, with $X$ and $\theta$ representing the input and parameters respectively. The training data is generated as:

{\footnotesize
\begin{equation}
[X_u|Y_u]=
\begin{pmatrix}
e_{1}^u & e_{2}^u &...& e_{L}^u|e_{L+1}^u\\
e_{2}^u & e_{3}^u &...& e_{L+1}^u|e_{L+2}^u\\
...& ...& ... & ...\\
e_{t-L}^u & e_{t-L+1}^u &...& e_{t-1}^u|e_{t}^u\\
\end{pmatrix}
\end{equation}
}
For each sequence in training set, at time slot t, we have: $X=\{e_{t-L}^u, e_{t-L+1}^u,..., e_{t-1}^u\}$ and $Y = e_{t}^u$.

Security events' features are fed to an RNN layer which will operate to predict poisoning attack on the device for the next time step, while Softmax function is applied for the probability distribution. For training, mean square as the loss function defines the deviation between actual and predicted values, while Adam optimization minimizes the loss values.     
\vspace{-0.1 cm}
\subsection{TOPOLOGY MUTATION OPTIMIZATION}

Let $tp=\{x_1(\tau)...x_N(\tau)\}$ be the topology of the FL at iteration $\tau$, defined by participation of devices. Here, $x_u(\tau)$ is the variable indicating the participation of device $u$ in FL at iteration $\tau$ (1 for participation and 0 for not-being active). The objective is to implement the MTD scheme while preserving the required confidence for attack mitigation and optimizing the FL performance criteria. 

Since the devices evolve their learning model within iterations of FL, the time that takes a device receive new updates, and perform the recognition is recognition time. The objective function i.e., (2) optimizes accumulated loss and time experienced by devices. $F_u$ and $T_{Int}$ are the loss and recognition time experienced at device $u$, respectively. $\alpha$, $\beta$ defines the priority of loss and time in the optimization. 

\vspace{-0.1cm}
{\footnotesize
\begin{equation}
\begin{aligned}
\min\limits_{tp(\tau)} \sum_{\tau, u} \alpha.F_u(\tau) + \beta.T_{Int}[\tau](u),
\end{aligned}
\end{equation} 
}

\vspace{-0.1cm}
The optimization should be performed according to the anticipation profile for Byzantine/benign status of devices, represented by $P=\{p_1,...p_N\}$. Constraint (3) identifies the anticipation profile. Here, $\phi(X_u; \theta_u)$ is the anticipated poisoning probability for the device obtained by RNN.   

\vspace{-0.1cm}
{\footnotesize
\begin{equation}
\begin{aligned}
\forall u: p_u(\tau)= \phi(X_u; \theta_u).
\end{aligned}
\end{equation} 
}

\vspace{-0.1cm}
Constraint (4) ensures a confidence for attack mitigation. Here, $1.(B)$ is 1 if boolean $B$ is true, otherwize it is 0. By this constraint, the devices that have been predicted to be Byzantine with a high confidence of $C_H$, will be excluded in the training phase at iteration $\tau$. 

\vspace{-0.1cm}
{\footnotesize
\begin{equation}
\begin{aligned}
\forall u: (1 - 1.(p_u(\tau) \geq C_H)).x_u(\tau) = x_u(\tau).
\end{aligned}
\end{equation} 
}
\textbf{Time calculation.} 
The recognition time in one iteration of FL, for a device is the time to take the new parameters and inference for the input sample. As calculated by (5), it includes: a) the time slots for local training, b) the up-link parameter transmission and the aggregation $T_{ag}$, c) the down-link parameter transmission $T_{down}$. d) The inference time.   

\vspace{-0.1cm}
{\footnotesize
\begin{equation}
    \begin{aligned}
        T_{Int}^c(u) =  \max_{u}K. \frac{|D_u|.f^{cmp}_s}{f^{cmp}_u} + T_{ag} + T_{down}(u) + \frac{f^{inf}_u}{f^{cmp}_u},
        \end{aligned}
\end{equation}
}
In (5), $K$ is the local training iterations, $f^{cmp}_s$ and $f^{inf}_u$ are the number of CPU cycles to train unit of sample and to recognize for a given input using the learning model, respectively. In the rest, we explain the components involved in (5). 

The time for partial aggregation at a BS includes the parameters transmission time, and the aggregation operation. Eq. (6) is the calculation. Here, $|w_g|$ is the number of global weights and $f^{cmp}_w$ is the number of required CPU cycles to aggregate one unit of data.  

\vspace{-0.1cm}
{\footnotesize
    \begin{equation}
    \begin{aligned}
        T_{ag}^i = \max\limits_{u \in R_i} \frac{|w_g|}{R_{u,i}} +        \frac{|R_i|.|w_g|.f^{cmp}_w}{f^{cmp}_i},
    \end{aligned}
    \end{equation}
}
The aggregation time includes the time for partial aggregation, the parameters transmission and final aggregation at the cloud: 

\vspace{-0.1cm}
{\footnotesize
    \begin{equation}
    \begin{aligned}
        T_{ag} = \max\limits_{i=1..M} (T_{ag}^i + \frac{|w_g|}{R_{i,c}}) + \frac{M.|w_g|.f^{cmp}_w}{f^{cmp}_c}.
    \end{aligned}
    \end{equation}
}    
 
Downloading the parameters takes time depending on the parameters size and the available transmission rates:

\vspace{-0.1cm}
{\footnotesize
    \begin{equation}
    \begin{aligned}
        T_{down}(u) = \frac{|w_g|}{R_{c,i}} +  \frac{|w_g|}{R_{i,u}},
    \end{aligned}
    \end{equation}
}
According to the 6G wireless communication model in \cite{kianpisheh2024collaborative}, \cite{fadlullah2020hcp}, \cite{lu2020low}, the transmission rate available for a device to communicate with the BS is estimated as follows:

\vspace{-0.1cm}
{\footnotesize
    \begin{equation}
        \begin{aligned}
            R_{u,i} = B_i\ln( 1 + \frac{Pt_u.g_u}{\eta} ),
        \end{aligned}
    \end{equation}} 
 \vspace{-0.4cm}
{\footnotesize   
    \begin{equation}
        \begin{aligned}
            g_u = C_g.d_{u,i}^{-\alpha},
        \end{aligned}
    \end{equation} 
}
where $Pt_u$, $g_u$, $\eta$ are respectively, transmission power of the device, channel gain and background noise power. The channel gain which depends on the path loss fading coefficient i.e., $C_g$, distance between device $u$ and base station $i$ i.e., $d_{u,i}$, and path loss exponent i.e., $\alpha$, is given in (10). 

\subsection{Deep Reinforcement Learning for Topology Mutation}
The search space order is exponential with dynamicity in channel communication and malicious behaviors. MDP and RL has been used in the network applications to solve the optimization in an adaptive manner and under dynamic situations. We employ them due to: (i) Function (2) can be explained as the sum of loss and time values, in the current interval and the function value in the previous interval; thereby having the memoryless property (ii) Considering the parameters determining the current state e.g., malicious profile of devices, devices participation in FL, transmission rates, every action that is performed by the agent ends to a new state transition which depends on the current state. (iii) The function (2) is in the form of accumulated rewards. Through an iterative process of state observation, action performing and feedback receiving, the Q-values are estimated by Bellman equation \cite{mnih2015human}. Training based on observing all states/actions is impractical for the high dimension of the states/actions and dynamic state transitions, thereby inefficiency of conventional RL. To solve the issue, we adapt DRL \cite{mnih2015human}, that generalizes experienced states/actions to non-observed ones through a neural network-based approximation of Q-values.

\noindent \textbf{MDP Elements:} MDP Elements include:\\
\textit{State}: The features of the state of network at time $\tau$, are:
\begin{itemize}
    \vspace{-0.1cm}
    \item The transmission rates represented with matrix $\mathbb{R}_{d,e}(\tau)$, at which the entry at row $i$ and column $j$ is the transmission rate between device $i$ and BS $j$, at iteration $\tau$. The variation in available bandwidth, location of devices, and channel gain causes dynamicity of the transmission rates. 
    \item Available CPU cycle at BSs and devices, represented by vector of $\mathbb{F}_{e}(\tau)$, and devices $\mathbb{F}_{d}(\tau)$ respectively.
    \item The current FL topology denoted by vector $\mathbb{T}\mathbb{P}(\tau)$ at which each entry indicates a device participation in FL.  
    \item The anticipated malicious behavior profile of devices $\mathbb{P}(\tau)$, where each entry specified the probability that a device will be Byzantine in the next iteration. 
\end{itemize}
\textit{Actions:} The action is decision about the FL topology mutation. The topology is mutated by defining the participants in FL for the next iteration (values of $x_u({\tau})$).\\
\textit{Reward:} To ensure constraint 4, if any of the devices which have been anticipated to be Byzantine with high confidence of $C_H$, be selected as a participant, the reward is 0. Otherwise, to optimizing the objective function, the reward is calculated as inverse proportion of accumulated of loss and time experiences at participants, as below:

\vspace{-0.3cm}
{\footnotesize
\begin{equation}
\begin{aligned}
\mathbb{R}(s(\tau),a(\tau))=\frac{1}{\sum_{u} \alpha.F_u(\tau) + \beta.T_{Int}[\tau](u)}
\end{aligned}
\end{equation} 
}

\begin{figure}[t]
\begin{flushleft}
    \includegraphics[width=5.4cm, height=3.2cm]{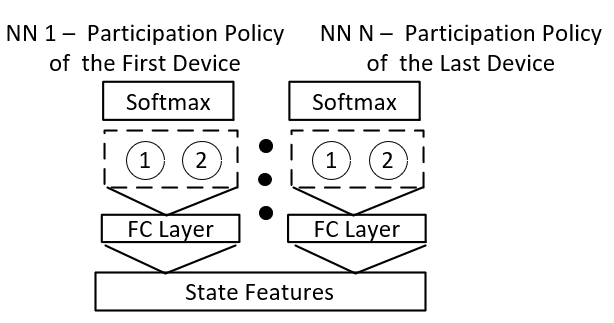}
    \label{fig:acc}
    \includegraphics[width=2.9cm, height=3.2 cm]{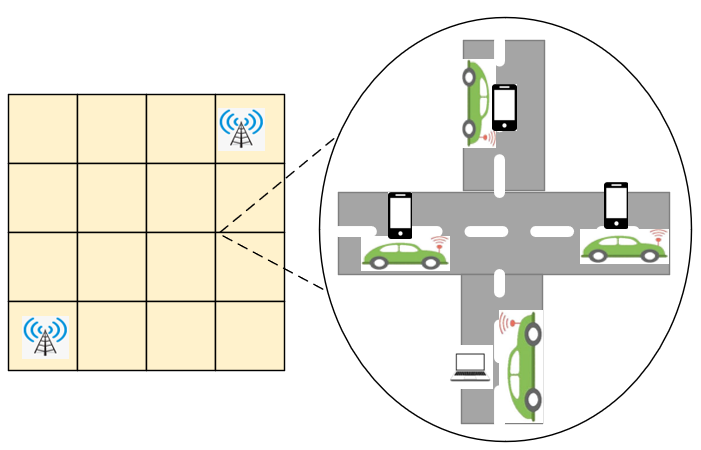}
     \label{fig:cost}
    \caption{(a) Policy networks. (b) Simulation grid.}   
    \vspace{-0.8cm}
\label{fig:NN}
\end{flushleft}
\end{figure}

\noindent \textbf{Training:} Using $N$ policy networks, the decision policy is derived by training them. The Neural Networks (NNs) represent the topology mutation policy, such that each NN specifies the participation of a device in FL. The input neurons are the state features. There is a Fully-Connected (FC) layer, with Softmax activation function. There are two neurons at the output layer. The first neuron in the output layer of NN $u$, indicates the Q-value for the participation of the device $u$ in the next iteration. The second neuron indicates the Q-value of the action of not-participation of the device in federation. (See Fig. 2.a).
Each episode consists of a run of FL. There are variation in security-related and network-related parameters at each FL iteration. Arrival traffic at devices varies and will be malicious in the case that a device is under attack. Network parameters e.g.,  network communication status, location of devices, and compromised devices change within iterations to reflect dynamicity in attack and network. Training is done through two steps performed at every iteration: 
\begin{itemize}
    \item \textit{Topology Mutation Exploration:} According to $\epsilon-$greedy policy, with a $1-\epsilon$ probability a device will either participate or does not participate, randomly (according to a uniform distribution). Otherwise, the current state features is given as input to the NNs. After neural operations, the highest Q-value at output layer defines the participation. If the highest value is for the first neuron, the device will participate. Otherwise, it will not participate. 
    
    \item \textit{Updating the weights:} After the topolody mutation exploration, reward is calculated, accordingly the NNs' weights are updated by Gradient Descent (GD) method, and using Bellman equation \cite{mnih2015human}.
\end{itemize}
The training can be performed offline and the topology mutation decision can be done in $o(N)$. 


\section{Experimental Results}
Simulation has been done by Python using an Intel(R) Xeon(R) Platinum 8358 CPU with 2.6 GHz frequency and 64 GB RAM. TensorFlow and Keras have been used to implement the deep learning. 10 devices with random CPU frequency in the range of 1.9 up to 2.4 GHz and transmission power of 23 db \cite{lu2020low} are moving by vehicles and perform FL for a DDoS attack detection. Recent studies e.g., \cite{kianpisheh2024collaborative}, \cite{li2021fleam} have shown FL can enhance accuracy of DDoS attack detection through sharing detection models, in comparison with individual learning approaches. 

There is a a $4 \times4$ grid environment with $100~ m$ width for each cell, where grid lines are bidirectional roads (Fig. 2.b). We have used SUMO simulator~\cite{SUMO} for generating mobility traces of vehicles. Mobility traces have been generated according to Manhattan model in urban areas~\cite{SUMO} with the mean mobility speed of 45 km/h, probability of 0.5 for go-straight and 0.25 for turn-right/lef at conjunctions. 

The learning is performed in the level of edge. In the locations $[50, 50], [350, 350]$, two BSs with coverage radius of 300 m, provide MEC with CPU frequencies of 3.2 and 2.6 GHz \cite{lu2020low}. The transmission bandwidth of BSs are 28 and 30 MHz \cite{kianpisheh2024collaborative}. The transmission power of BSs is 34 db. Path loss exponent is 5 and background noise power is -174 db.m \cite{lu2020low}.    

We used CICDDoS 2019 data set~\cite{sharafaldin2019developing} comprising legitimate traffic and traffic for DDoS attacks of UDPLag and SYN DDoS attacks, through protocols of HTTP, FTP. The dataset provides 87 IP flow features e.g., source/destination IP addresses/ports, protocols, flow packet statistics, flag-related information etc., based on which the attack is detected. At each iteration, in the range of 2000 to 10000 instances are randomly distributed among devices for training. 2500 random instances are distributed for test. We found GRU with hidden layer size of 8 neurons and Adam optimization efficient for attack detection \cite{kianpisheh2024collaborative}. The feature matrix for the packets in a flow forms rows of patterns which is given as input to the GRU. The out-layer predicts the occurrence probabilities of packets as a function of previous observations. A flow (with size of 10 packets) is malicious if the ratio of the malicious packets is higher than a threshold 0.7. We discussed the details and efficiency of the model parameters in \cite{kianpisheh2024collaborative}.

FL is applied in a system without attack (FL) and a system with poisoning attack. At each episode, the adversary targets 2 to 4 random devices and sends malicious traffic to the compromised devices to take control. We used the Botnet attack trace in \cite{sharafaldin2018toward} to generate the malicious traffic. It has recorded the legitimate and attack traffic features (total of 78 features) e.g., traffic duration, total packets, packet/flow/header/segment statistics etc. The Botnet attack provides capabilities for the attacker to attack devices and do e.g., remote sell, file upload/download, key logging \cite{sharafaldin2018toward}, thereby enabling taking control for intervening in FL process. In this regard, a traffic flow i.e., 11 sequences from the Botnet trace which illustrate an attack occurrence at the end, are sent to a compromised device ($L = 10$). Then, the attacker emulates malicious local models for updating. Two attacks have been simulated:
\begin{itemize}
    \item \textit{Attack 1:} The attack in \cite{al2023untargeted} has been adopted. The attacker tries to deviate from global model with an arbitrary malicious model. Poisoned model is calculated by updating global model with a learning rate based on the difference between the malicious target model and the \textit{global model}. To maximize the effect of poisoning, we set learning rate as 1 and the target model parameters as $-10$ times of global model parameters. We also added a small random noise in poisoning process.
    \item Attack 2: It is adopted from attack in \cite{baruch2019little}. Using a Gaussian-based statistics, it deviates each dimension of the model parameter from the \textit{mean} with a fraction of the \textit{standard deviation}. \cite{baruch2019little} gives the mathematical details (See \cite{baruch2019little} for the fraction calculation).    
\end{itemize}

For attacks implementation, \textit{global model} is estimated by averaging the weights of devices and \textit{mean} and \textit{standard deviation} are calculated over the devices' models. The DRL process has more explorations in earlier episodes, and the exploitation gradually increases up to the greedy selection of 98\% in the last episode. Discount rate of 0.1 and ADAM optimization in DRL operated efficiently. Channel conditions (due to devices mobility), the compromised devices varies in iterations.  $C_H$ is 0.75. The results are average of 25 runs. 

{\footnotesize
    \begin{center}
    \vspace{-0.3cm}
    \captionof{table}{The results of Botnet attack anticipation.}
    \vspace{-0.2cm}
     \begin{tabular}{|p{2 cm}|p{1.3cm}|p{2cm}|p{2 cm}|}  
         \hline
         \textbf{Model}
         & \textbf{Accuracy}
         & \textbf{FP}
         & \textbf{FN}\\
         \hline
        GRU 5 & 0.87 & 0.06 & 0.92\\         
         \textbf{GRU 8} & \textbf{0.77} & \textbf{0.24} & \textbf{0.27}\\         
         GRU 11 & 0.51 & 0.49 & 0.37\\
         LSTM 8 & 0.70 & 0.26 & 0.76\\
         \textbf{LSTM 11} & \textbf{0.88} & \textbf{0.06} & \textbf{0.88}\\
         LSTM 16 & 0.71 & 0.26 & 0.62\\
         \hline
    \end{tabular}
    \label{tab:bot}
    \end{center}
}
\vspace{-0.2cm}
We exploited 12000 and 30000 random sequences of benign/attack events in the Botnet trace, in the training and the test phases, respectively. Table \ref{tab:bot} shows the performance. We evaluated GRU and LSTM with 5, 8, 11, 16 hidden neurons. LSTM 11 has achieved the highest accuracy of 88\% and a False Positive (FP) of 0.06. However its False Negative (FN) is 88\% which is high. GRU with 8 hidden neurons has gained accuracy of 77\% and a lower FN of 0.27 and a FP of 0.24. Indeed, it recongnizes attacks and benign traffic with precision of 73\% and 76\%, respectively. As the nature of attacks in \cite{sharafaldin2018toward} is complex, we could not get a better performance. However, this performance is promising in the DDoS attack detection.     
\begin{figure}[t]
\begin{center}    \includegraphics[width=5.4cm, height=4cm]{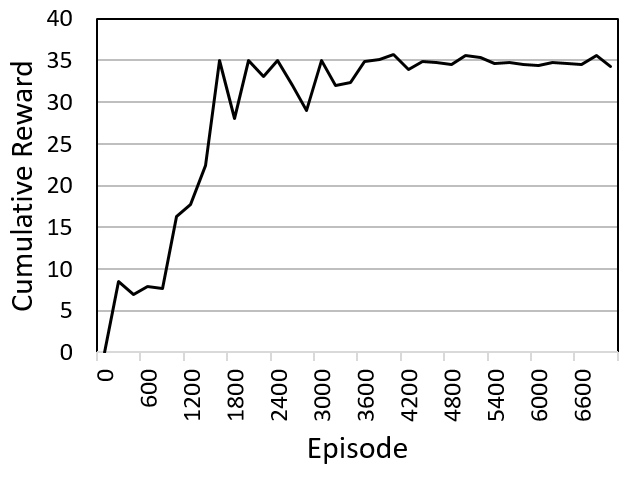}
    \label{fig:acc}      
    \vspace{-0.3cm}
    \caption{Reward variation during training.}   
    \vspace{-1cm}
\label{fig:reward}
\end{center}
\end{figure}

Fig. \ref{fig:reward} shows the cumulative reward gain within episodes in MTD-FL. The cumulative reward has increased up to 35 and become stable after episode 3600, which illustrates the evolution of the topology mutation policy and convergence. 

\begin{figure*}[!h]
\begin{center}
    \begin{minipage}{5.5cm}
    \begin{center}
        \includegraphics[width=5.8cm, height=4cm]{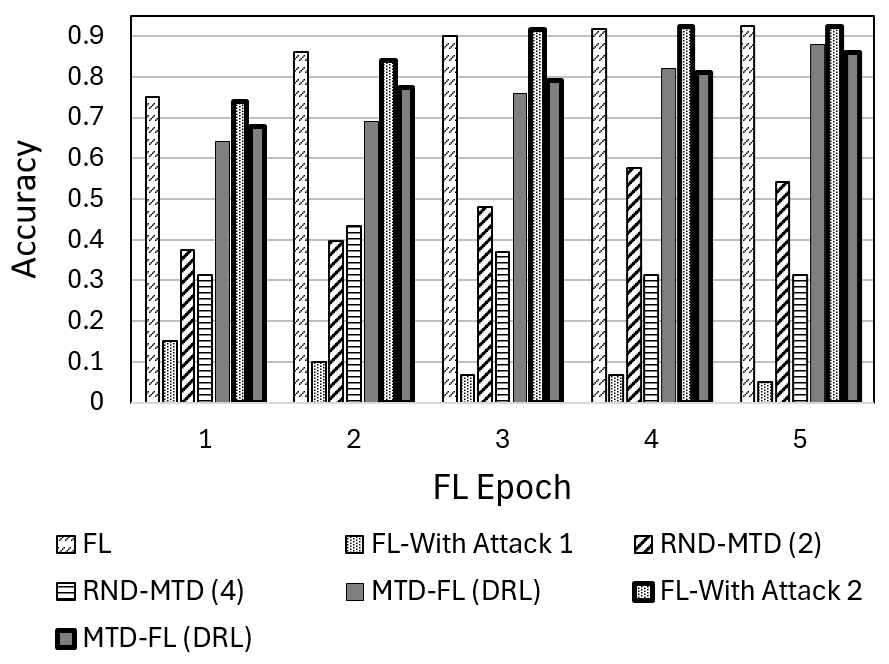}
    \end{center}
    \end{minipage}
    \begin{minipage}{5.8cm}
    \begin{center}
        \includegraphics[width=5.5cm, height=4cm]{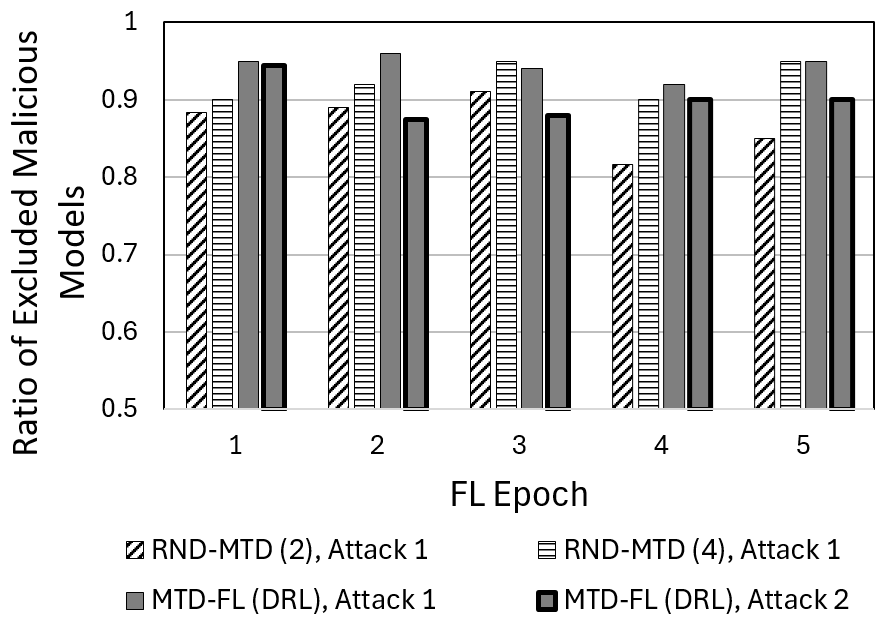}
        \end{center}
    \end{minipage}
    \begin{minipage}{5.8cm}
    \begin{center}
        \includegraphics[width=5.4cm, height=4cm]{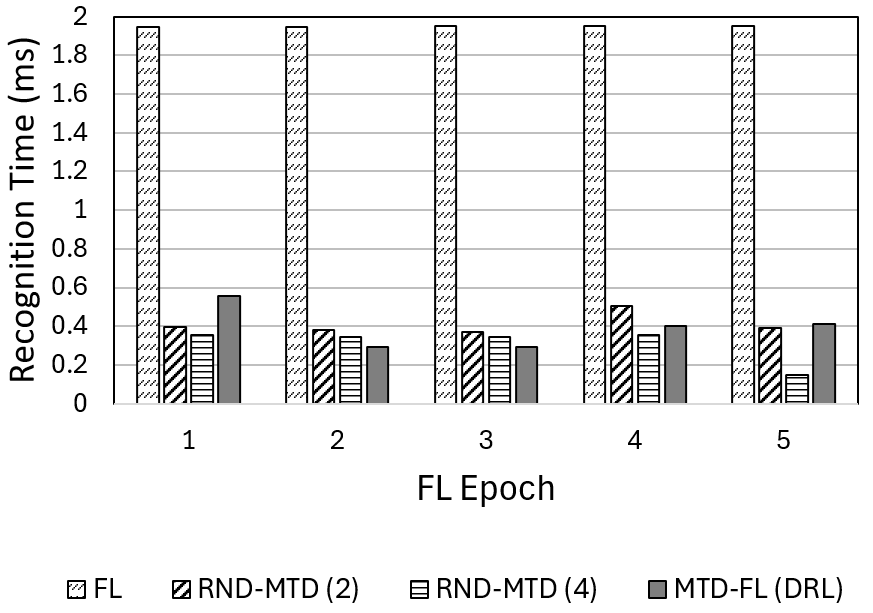}
        
    \end{center}
    \end{minipage}
    \label{fig:P}
     \caption{Comparison of MTD-FL vs. baselines. a) accuracy, b) ratio of excluded malicious models, c) recognition time. }
\end{center}
\vspace{-0.8cm}
\end{figure*}
Fig. 4. compares MTD-FL performance with baselines: i) \textit{FL} is the scenario of FL without attack. ii) \textit{FL-With Attack} which is the scenario after poisoning attack. iii) \textit{RND-MTD} where topology mutation happens by blocking 2 or 4 random participants  at each iteration. The bars with bold boarders are for Attack 2 scenarios, and the rest are for Attack 1 scenarios. 

Fig. 4.a illustrates the accuracy of DDoS attack detection. Due to federation, the accuracy of attack detection increases from 75\% in the first iteration to 92\% in the fifth iteration. After Attack 1, when there is no MTD strategy, the accuracy decreases to 15\% in the first iteration. The side effect worsens in iterations, reaching  5\% in the last iteration, due to more injection of poisoned models. RND-MTD with blocking 2 devices to form FL topology, has a chance to exclude malicious devices and increases the accuracy up to 57\%. However the accuracy reduces when it blocks 4 devices in federation. The reason is that it probably excludes more benign models due to random topology mutation policy. For attack 1, MTD-FL raises the accuracy up to 88\%. The reason is that it recognizes the malicious models by traffic analysis and excludes them in the federation through a DRL-based optimization. In early iteration there is accuracy loss in comparison with FL without attack. The loss is mainly due to FP (See Table I) and possible blocking of benign models. However, the accuracy loss is rather compensated in later iterations when more models are federated. For Attack 2 scenario, even after poisoning the accuracy is competitive with the case that there is no attack. Since, as discussed in \cite{baruch2019little}, this attack manipulates the distribution in bounds of other distributions of devices. MTD-FL gains accuracy up to 86\%. We envision with lower FP accuracy can be competitive with scenario without attack.  

Fig. 4.b shows the ratio of excluded malicious models. Generally, for Attack 1, MTD-FL has outperformed RND-MTD and could exclude 92\%-96\% of malicious models in the aggregation phase. The reason is that in contrast with random topology mutation strategy in RND-MTD, in MTD-FL the topology mutation is optimized based on anticipation profile for Byzantine/benign status of devices, thereby omitting the Byzantine anticipated models to maximize the reward. Under Attack 2, MTD-FL could exclude 87\%-94\% of malicious models. In \cite{baruch2019little}, Attack 2 has shown to be unrecognizable by most outlier detection methods. Generally, even nontrivial detectable attacks like Attack 2, still have the potential for global model manipulation, particularly by targeted and smart-crafted Byzantine parameters. MTD-FL through traffic analysis can detect malicious model, thereby promising to reduce the attacker intervening chance in the learning process. 

Fig. 4.c illustrates the recognition time. The recognition time of FL is around in the range of 1.94 to 1.95 ms. In RND-MTD (2), 8 devices participate in learning, and less time will be spent on trainig and parameter transmission thereby, lower recognition times are experienced. Time becomes slightly lower in RND-MTD (4) due to the participation of 6 devices in federation. MTD-FL has reduced recognition time by 1.6 ms in comparison with FL by considering time in optimization and saving time slots required for training and parameters transmission of devices that are anticipated to be malicious.

\vspace{-0.04cm}
\section{Conclusion} \label{sec:sec5}
This paper exploits device-level traffic analysis to anticipate the Byzantine status of devices and provides a MTD-based FL that empowers FL against model poisoning attack. RNN-based mechanism for traffic analysis and establishing security profile of devices has been given. Optimization framework for MTD strategy and a deep reinforcement mechanism with capability of adaption with malicious activities and wireless communication status have been provided. The method has been evaluated with two attacks. Simulation results for a DDoS attack detection scenario, illustrate reduction of recognition time and improvement in accuracy.  This paper investigates poisoning under Botnet attack.  Assessing the performance of the method under other attacks is a future work. 

\vspace{-0.08cm}
\section*{Acknowledgment}
This work was supported in part by the European Union’s Horizon program through the RIGOUROUS project (Grant No. 101095933) and 6G-SANDBOX project (Grant No. 101096328). The paper reflects only the authors’ views. The Commission is not responsible for any use that may be made of the information this paper contains.


\end{document}